# Epistemic-Pragmatist Interpretations of Quantum Mechanics:

# A Comparative Assessment

Ali Barzegar & Daniele Oriti[1]

Arnold Sommerfeld Center for Theoretical Physics and Munich Center for Mathematical Philosophy, Ludwig-Maximilians-University, Munich

## Abstract

In this paper, we investigate similarities and differences between the main neo-Copenhagen (or "epistemic-pragmatist") interpretations of quantum mechanics, here identified as those defined by the rejection of an ontological nature of the quantum states and the simultaneous avoidance of hidden variables, while maintaining the quantum formalism unchanged. We argue that there is a single general interpretive framework in which the core claims that the various interpretations in the class are committed to, and which they emphasize to varying degrees, can be represented. We also identify, however, remaining differences of a more substantial nature, and we offer a first analysis of them. We also argue that these remaining differences cannot be resolved within the formalism of quantum mechanics itself and identify the more general philosophical considerations that can be used in order to break this interpretation underdetermination.

[1] . barzegar11@gmail.com, daniele.oriti@physik.lmu.de

## 1. Introduction

Quantum Mechanics (QM) is the scientific framework used to formulate basically all our best-confirmed theories for physical phenomena (with the notable exception of gravitational physics). Despite this, there is no consensus on its interpretation yet. That is, there is no consensus on what, fundamentally, QM says about the way the world is, in its basic principles, and on what exactly, in the QM formalism, refers to it. This is the strong notion of 'interpretation'. One can instead have a mild realist position where 'the properties of the world' are inferred from the theory, in the sense of asking which world would be compatible with the success of the theory as a shared tool for navigating experience, without turning this into a statement about how the world actually is. One can also adopt an even weaker notion of interpretation: one is 'interpreting' a theory if it is trying to 'find out what the theory is about, including the possibility that the theory should be understood purely as a convenient mathematical tool for prediction'. From this point of view, also a strong instrumentalist (or anti-realist) position would amount to an interpretation. One could use any of these notions of interpretation to read our analysis and its results, and our analysis and results would remain valid and interesting in the same way.

In any case, a number of things that QM *seems to say* about the world, when taken at face value, are at odds with common sense and with the received (and by now almost common-sensical) view from classical mechanics in its conceptual foundations.[2] By this, we do not just mean that QM makes factual statements about the world that go against what common sense would have made. This is what happens almost inevitably with any theory in fundamental physics, and it is to some extent part of the very nature of frontier science. What we mean is that QM appears to shake the very conceptual and ontological foundations of our classical picture of the natural world.

---

[2] . By the common-sensical character of classical physics we mean the presupposition that objects possess their properties intrinsically at every moment of their history. Moreover, these properties have definite values at all times independent of our interventions.

The conceptual issues raised by QM are many and multi-faceted (e.g. see Albert 1994; Lewis 2016; Myrvold 2018), and existing interpretations can be classified according to their proposed solutions to such issues.

The nature of the quantum state ("the wavefunction") is a key issue. There are two main classes of interpretative approaches to QM: ontic and epistemic interpretations. According to the ontic interpretations, the quantum state is part of the furniture of the world, in direct correspondence with elements of reality, thus a real physical quantity in itself. The many-worlds interpretation, the de Broglie-Bohm theory, and spontaneous collapse models are popular examples of the ontic camp. On the other hand, according to the epistemic interpretations, the quantum state encodes the knowledge, information or beliefs that an "observing" system has about (its past/future interactions with) the "observed" system and is a "tool" for predicting the results of future interactions with it[3].

Within the epistemic camp a further distinction ensues depending on whether this epistemic attribution (summarized by the quantum state) refers to an underlying ("hidden") ontic structure of the physical system under consideration, which QM models or reflects only partially, but that represents the actual reality of it. A view like this would match standard scientific realism according to which scientific theories 'discover and map out an already structured and mind-independent world' (Psillos 1999; xvii), somehow saving it from the severe challenge of QM, which would then "only" challenge our view of how we access such world; in other words, the QM challenge would be only confined to the epistemological, rather than the ontological level, where the ontological level will then depend on how we model the "hidden structure". Epistemic interpretations of this kind are also dubbed "hidden variables models". It should be noted that there is a difference between psi-epistemic and psi-ontic hidden variable interpretations. As an example, Spekkens' toy model is a psi-epistemic hidden variable interpretation in which the epistemic quantum state is defined over a set of ontic hidden variables (Spekkens 2007). On the other hand, Bohmian

[3] We will discuss in the following how the word "epistemic" in this context should be used, if at all, with much care, because it may seem to imply directly some underlying ontic structure one has knowledge about. This is only the case for a subset of "epistemic" interpretations, and it leads to a misunderstanding of others.

Mechanics is a psi-ontic hidden variable theory in which the ontic quantum state determines the particle trajectories (Goldstein 2024).

In this work we only concern with those epistemic interpretations which posit no hidden variables. We refer to them as "epistemic-pragmatist interpretations"[4]. According to them, the quantum state is a complete characterization of the physical system and we should refrain from postulating any hidden ontic structure or an “observer”-independent state of affairs.[5] QM (and possibly physics more generally) deals only with the perspective of a particular system when it enters into relationship with other systems. As we shall see, this means that pragmatist epistemic interpretations naturally adopt some more nuanced (in particular, perspectival or relational) form of realism. Notable interpretations in this (broad and diverse) camp include QBism (see Fuchs (2002, 2010)); Relational QM (see Rovelli (1996)); information-theoretic perspectives (see Zeilinger (1999, 2005), Mueller (2017); Brukner (2017), Bub and Pitowsky (2010)), Healey’s pragmatism (2012, 2017); and more (see for example Friederich (2015) and Boge (2018)).

In this paper[6], which of course takes on board the results of much earlier work, we investigate the similarities and differences between the pragmatist epistemic interpretations of QM. In section 2, we identify a common core which characterizes the epistemic-pragmatist perspective as a whole and is shared, we argue, by all these different epistemic interpretations. In this sense, they can be regarded as different instances (with different emphasis on or specific declinations of the same characterizing features) of a single general interpretative framework. However, as it will be argued in section 3, there remain important differences between these epistemic-pragmatist interpretations; they concern mostly the nature of the observing system relative to which the quantum state is defined and the interpretation of quantum probabilities. In the same section 3, we discuss also if and how this interpretation underdetermination can be resolved within the QM formalism and which other kind of analysis may help resolving it. Section 4 summarizes our conclusions.

---

[4] For the reasons mentioned in the previous footnote and to be discussed later, one could also simply call this class of QM interpretations "pragmatist", dropping the epistemic connotation, which only serves to distinguish them from the ontic interpretations of quantum states. With this in mind, we stick to the chosen label, for now.

[5] The reason for the quotation marks around “observer” will be clear in the following, but it has to do with the fact that the precise characterization of the second system with respect to which the quantum state of a physical system is defined is what distinguishes the various versions of the epistemic-pragmatist perspective on QM. In fact, partially for the same reasons, similar quotation marks could be added around "epistemic" in the characterization of these interpretations of QM, as we will discuss in the following.

[6] While this paper was being completed, an interesting analysis by Pienaar has appeared (Pienaar 2021) on the arXiv. Pienaar's analysis is limited to a comparison between RQM and QBism, but reaches several of our own conclusions, which comforts us about their correctness.

## 2. The common core of the epistemic-pragmatist interpretations of QM

In this section, we explore the common core which characterizes all the different epistemic-pragmatist interpretations of QM. We do by first articulating more precisely what are the key shared tenets constituting this core. Next, we discuss each of the main QM interpretations that, we argue, belong to the epistemic-pragmatist class, showing how they indeed share these tenets and the different ways in which they incorporate or express them. Here, we should note a caveat concerning the metaphysical element of "participatory realism" in our framework. On the one hand, the notion of "participatory realism" as a metaphysical position remains to be clarified and developed; on the other hand, none of the epistemic-pragmatist viewpoints have, in our view, spelt out a precise metaphysical position underlying their interpretation of QM, and in fact have expressed evolving views, so what they really think or what their interpretation supports remains ambiguous; and this of course, makes it harder to settle the question of whether one specific interpretation is compatible or in agreement with such metaphysical commitment. Thus, while we could try to treat these three elements as a shared common core, due to these difficulties and ambiguities on the metaphysical component, we propose them instead as three axes of an interpretative framework.

We note here that, in this conceptual framework and in the QM interpretations that, we argue, adopt it, it is a combination of 1 and 3 (thus, the adoption of a new, somewhat weaker notion of realism) that allow to make sense of the violation of Bell's inequalities without renouncing locality.

The first element of the epistemic-pragmatist framework for QM is the following:

1. An "epistemic" (as opposed to ontic) view of quantum states

A quantum state is not in itself an element of physical reality. It represents the knowledge, information or beliefs of the "observing system" when it enters into a relation with the "observed system", e.g. through a measurement. An important consequence of this epistemic reading of quantum states is that the wavefunction collapse following a measurement is interpreted, rather straightforwardly, as the update of the epistemic state of the observing system upon acquiring new information. As we had mentioned, this is the defining aspect of epistemic interpretations in contrast with ontic ones, like many-worlds, spontaneous collapse or de Broglie-Bohm.

Next, distinguish between the hidden variables epistemic interpretations and the pragmatist-epistemic interpretations. The former regard the quantum state as a statistical distribution on the hidden variables and so add new elements to the formalism of QM in order to make their interpretation work. All the latter, on the other hand, regard quantum mechanics as complete and therefore leave the formalism intact. At a technical level, this translates (also) in the fact that these latter interpretations can not be analyzed within the framework of ontological models, which has the distinction between ontic and quantum mechanical states at its core, and thus that important results, e.g. Psi-ontology theorems, do not apply directly to them (Leifer 2014; Ben-Menahem 2017).

Thus, another characterizing feature of epistemic-pragmatist interpretations is that they do not suggest any need for modifying QM (for the purpose of interpretation, at least). This is a technical issue tied to the mathematical foundations of QM notably Gleason and Bell theorems (see e.g. Demopoulos 2022). However, we could still imagine modified QM theories motivated by specific physical reasons that would still qualify as epistemic-pragmatist in terms of all the deeper conceptual issues that QM raises[7].

[7] Our point here is simply to stress that, while epistemic-pragmatist interpretations do not require or motivate specific modifications of the quantum formalism, they do not exclude them either. In fact, an epistemic-pragmatist interpretation can be (but certainly does not have to) applied equally well on most generalized probabilistic frameworks, due to their rather operational language (see Plavala (2021) for a general review, and Barrett (2007) for those putting information processing at the forefront).

The second distinguishing feature between the hidden variables and epistemic-pragmatist camps is more relevant. It concerns realism. Hidden variables approaches tend toward a traditional form of realism in terms of an absolute ontic structure that physical theories should represent from a God's-eye point of view. Here, the ontological commitments are prior to and define the epistemology. In contrast, the pragmatist approaches are willing to weaken traditional realism and, if they decide to commit to realism at all, try to infer the appropriate ontological commitments from the theoretical framework of QM only. In a way, this is a modern instance of the age-old issue concerning the respective role of metaphysics and epistemology, which Einstein nicely summarized as follows:

> The real difficulty lies therein that physics is a kind of metaphysics; physics describes "reality". But we do not know what "reality" is; we only know it through physical description!' (von Meyenn 2011, 537)

For the pragmatist epistemic interpretations, thus, epistemology has primacy over ontology. However, this is not all. This difference in attitude, so to speak, becomes a substantial difference in philosophical content, when applied to QM, in particular in the suggested ontology. The ensuing issue, indeed, is whether the "reality" indicated by the epistemology (concretely, in our case, the picture of reality we can infer from QM) remains an "observer-independent" one, whether it remains, like in traditional realism, about properties attributed to entities "in themselves", regardless of how such entities have been probed and their properties ascertained. In different specific ways, we argue, epistemic-pragmatist interpretations of QM deny that this is the case.

The second characterizing feature of the epistemic-pragmatist framework is:

2. A metaphysics of participatory realism

QM is about the *relation* between two pieces of the world, and this relation is an irreducible, constitutive component of reality and of any characterization of "physical systems". This is the key lesson that epistemic-pragmatist interpretations take from QM and turn into a foundation for metaphysics. This interaction is usually called a 'measurement' in quantum theory, but it does not have to correspond to measurements taking place in a lab as part of an experimental procedure and should be understood more generally.

In the corresponding metaphysical framework, then, interactions or relations should be regarded as the fundamental, primitive elements of reality. Therefore, what one regards as objects or substances (the "physical systems" themselves) should be defined in terms of the more fundamental relations. That is, we should move from an object-based ontology to a relation-based one. Since what is fundamentally real is only such relation, reality is in a specific sense *created* or at least *continuously shaped*, *moulded* by the interaction between the two involved physical systems, and at the moment in which such interaction takes place. This is a "participatory" and "relational" realism: we have an ontology of relations according to which physical systems acquire properties (and thus, in a sense, become real) only when they enter into relationships with each other[8].

So, according to participatory realism, the relations are created upon the interactions between the systems. There are no pre-existing relations or structures out there. In this picture, 'there is no such thing as the universe in any completed and waiting-to-be-discovered sense' (Fuchs in Schlosshauer 2011, 285). There is only 'a reality in the making' (Popper 1995, 20). Articulating the metaphysics of participatory realism is an important and challenging task, worth of an entire research programme, thus going beyond the more limited scope of this article. It amounts to identifying and clarifying different ways in which we participate in making reality, and correspondingly rewriting the categories and concepts of traditional metaphysics. The research programme would not have to start anew, and several insights can be found in both the recent philosophical literature (see e.g. Atmanspacher and Rickles 2022) and in older ideas (e.g. the works of W. James).

[8] Thus, the notion of an isolated system is just an abstraction also at the fundamental metaphysical level, not just at the physical one.

Thus, a certain reading of QM states and of the main elements of QM, together with an "epistemology-first" attitude, leads to a new metaphysics.

The third characterizing element of an epistemic-pragmatist framework is:

3. An epistemology of perspectival objectivity

This epistemological position is the natural consequence of the relational view on quantum states coupled to the conviction that they provide a complete account of physical systems (even if it is one that contains less information than that provided by classical physics, due to Heisenberg uncertainty). If quantum states are a complete account of physical facts and they are relational, thus always and only about the relation between the "observed" physical system and the "observing" one, it follows that physical facts are necessarily perspectival. This relativity of facts in QM goes back to the “Wigner’s friend'' scenario (Wigner 1961). In this scenario, Wigner’s friend is inside a laboratory and performs a measurement on a system which results in a definite outcome. Wigner himself is outside the lab and assigns an entangled state to the system and the friend which he can experimentally verify. So, we have two different descriptions of the situation inside the lab. The crucial point is that one can regard both descriptions as true objectively only on pain of contradiction. They are true only relative to the perspective of Wigner or his friend. Specifying a second "observing" system, thus providing a perspective, is required in order to give meaning to physical properties of the first "observed" system. There is no perspective-independent fact. In other words, facts (about physical systems) are irreducibly relative (to a perspective provided by other physical systems). Indeed, Wigner's friend scenarios are probably the best context in which one could analyze the differences between the epistemic-pragmatist approaches, and what notion of objectivity is compatible with or underlies them. It is clear, however, and it is our point, that they all share the rejection of a strong form of objectivity that is context-independent and in which intersubjectivity is complete.

This is as straightforward as it is radical. Does it leave us with any form of objectivity, in our understanding of the world as encoded in QM? The only possible form of objectivity that is compatible with an irreducibly perspectival knowledge is a weaker notion of (perspectival) objectivity, which amounts to constraints on the myriads of possible perspectival accounts.

The constraints are of two types. First, there is a requirement of internal consistency regarding the individual perspective of a given observer. Second, there is the requirement that we can map between the perspectives of two observers. How strong is the resulting notion of (perspectival) objectivity depends on the required properties of this map between perspectives. A strong version is granted if we require that there is an isomorphism between any two perspectives on the same physical system (the same portion of the world), thus if there is a mapping such that there are no element in any of the two perspectives which is not mapped into an element of the other, and if this is in fact a one-to-one correspondence (one may call this, rather, a form of intersubjectivity). Also, of course, the mapping should not result in inconsistencies upon translation between the two perspectives. An even stronger form of perspectival objectivity would correspond to the case in which there would be an invariant element (or more) in the mapping, i.e. an element (or more) which is left identical when seen from both perspectives. This would then be regarded as an element of reality in the traditional "objectivist" sense of standard realism. For example, in the theory of relativity, there are Lorentz transformations which allow for mapping between different perspectives and so make possible a strong form of objectivity (or intersubjectivity). Moreover, the spacetime interval is an invariant element of the theory which ensures a stronger form of objectivity within the theory.

Traditional objectivity of standard realism, amounting to an entirely "observer-independent" reality, would correspond to the limiting case in which all elements of the different perspectives are invariant under mapping. The existence of an isomorphism is all that is required to have the strongest form of perspectival objectivity (i.e. a perfect intersubjectivity) allowed by a relational view on quantum states and a metaphysics of relations. A one-to-one correspondence between my perspective and yours, so that I can translate your perspective into my own and vice versa, i.e. an exact covariance of perspectives, and the possibility for me to know (by application of the translation map) what your perspective on the physical world is, thus resulting in intersubjective agreement, suffices for perspectival objectivity.

Let us also touch upon perspectivism in the philosophy of science literature. According to Giere (2006), scientific knowledge both in its observational and theoretical aspects is perspectival. And as Massimi (2018) emphasizes, the perspectival nature of scientific knowledge is incompatible with the objective realist's supposition of 'One True Theory' towards which science is progressing. Scientists always face a plurality of models, theories and experimental methods. Then, how does perspectivism manage to avoid relativism? The strategy to secure some form of objectivity and avoid radical relativism is by relying on the existence of isomorphisms between perspectives, when there are no invariant elements across different perspectives. These isomorphisms guarantee a form of weaker objectivity, while invariant elements would be our best candidates for what we could regard as objectively real in the strongest, traditional sense. We should do science upon the belief that our best invariant elements will continue to hold in the alternative perspectives to be discovered. Although there is no guarantee that this program will succeed forever this is the best we can do.

This concludes our proposal for the basic ingredients of an epistemic-pragmatist interpretative framework for QM, that we argue brings together a number of existing specific interpretations. However, we do not claim beforehand that all the different interpretations are committed to the core elements of this framework to the same extent and degree. In fact, it may also turn out, after considering the differences between the various interpretations in this class, that one or more of them, in fact, should be excluded from the framework, since the differences turn out to be too profound and substantial. But this does not mean that the whole framework does not exist and does not comprise at least some of them; and it does not make the characterizing features we identify less interesting.

In the following subsections, we discuss the different implementation of these basic ingredients within each of the QM interpretations we suggest to understand in a unified manner. The upshot of our analysis will hopefully be to convince the reader that there is indeed a single general epistemic-pragmatist framework, characterized by the common core identified above.

### 2.1. Bohr's view

The above elements can be already detected (often with some effort) in the positions expressed by Bohr during the laborious birth of QM. This is the reason why the QM interpretations we discuss in some more detail below are often grouped under the label of "neo-Copenhagen" (see Cabello, 2015).

Here we could just give an outline of Bohr's interpretation of QM. Discussion of the nuances of his philosophy of QM is beyond the scope of this paper.[9] The following quote is a good illustration of Bohr's interpretative approach to QM:

> The entire formalism is to be considered as a tool for deriving predictions, of definite or statistical character, as regards information obtainable under experimental conditions described in classical terms and specified by means of parameters entering into the algebraic or differential equations of which the matrices or the wave functions, respectively, are solutions. (Bohr 1950, 52)

First of all, for Bohr, the quantum state has an epistemic or instrumental nature. It refers to our information about the system under study (or the one that can be extracted by observations/interactions with it) rather than a description of its intrinsic properties; it is not in itself an intrinsic property of the physical system. Second, any application of QM requires a description of the whole experimental arrangement, which defines the context of our interaction with the system and for attributing it certain properties upon observation/measurement. Third, this description of the context should be formulated in classical language, i.e. using the terms and concepts of classical mechanics. This is, for Bohr, a logical requirement which enables us to communicate the experiment and its results to others. Fourth, Bohr calls the results of the observations of a given physical system

[9] . See e.g. Faye and Folse (2017); Zinkernagel (2016).

obtained under specified experimental conditions a ‘phenomenon’. QM is a theory about (our knowledge of) phenomena. A phenomenon is inseparable from the experimental arrangement under which it could appear, just as it is inseparable from (since it refers to) the observed physical system. A phenomenon is the pair of observed system and context of observation. In other words, one cannot consider properties of the object under study as intrinsic, thus existing independent of the experimental conditions in which they are ascertained. A phenomenon is a holistic or composite entity. This holistic nature of quantum phenomena leads to the idea of *complementarity*. In classical physics, all the properties of the object can in principle be ascertained by a single experimental arrangement or several ones, but these arrangements do not need to be specified. However, in QM, properties obtained from different experimental arrangements exhibit complementarity, in that not all of them can be simultaneously realized or ascertained. Still, though their combination into a single unified picture appears contradictory, they collectively exhaust all the conceivable knowledge about the object (Bohr 1963).

Thus, notwithstanding the indivisibility of the quantum phenomena, one should make a classical-quantum division within one’s theoretical description of the experimental arrangement. Of course, for Bohr, everything is ultimately describable quantum-mechanically. What makes the classical-quantum division possible, is a difference of scale between the classical measuring apparatus and the quantum object under study. Usually, the measuring apparatus operates at a scale where the quantum effects could be neglected. According to Bohr (1935), this classical-quantum division is ‘largely a matter of convenience’. It is not an ontic but an epistemic distinction which allows for the experiment to be described in classical terms. It is only under such distinction however that physical systems can be said to "possess" specific properties, in the context-dependent (thus, relational) sense specified above.

Lastly, let us emphasize the participatory nature of quantum reality in Bohr’s view:

> Now the quantum postulate implies that any observation of atomic phenomena will involve an interaction with the agency of observation

> not to be neglected. Accordingly, an independent reality in the ordinary physical sense can neither be ascribed to the phenomena nor to the agencies of observation. (Bohr 1928, 580)

That is, a phenomenon is the manifestation of an interaction between two parts of the world. And as Bohr (1950) emphasizes, we are 'both actors and spectators in the drama of existence'. We classify Bohr's position among the epistemic-pragmatist interpretations mainly because of his epistemological and methodological views. Indeed, it seems difficult to attribute to Bohr an uncontroversial metaphysical position. He is regarded as a realist by some authors and as an anti-realist or instrumentalist by others. We would maintain, though, that if one wants to attach a metaphysical position to the epistemological one and also avoid any hidden-variable type views, then some form of participatory realism must be invoked. Obviously, the specific form may differ in many ways between Bohr's and the others.

Here, it is worth mentioning another interpretive approach called "Contexts, Systems and Modalities" (CSM) (Auffèves and Grangier, 2016). According to CSM, the quantum state is defined only relative to a context and this context is always classical. This provides a form of contextual objectivity in the sense that the quantum states are associated with 'predictions that are certain and observer-independent in a given context' (Grangier 2005, 18). It is clear that CSM is in line with Bohr's approach to QM, to the point that it can be seen as a formalization of it.

### 2.2. The Bub-Pitowsky Interpretation

This interpretation

> is mathematically expressed by taking the Hilbert space, or more precisely, the lattice of its closed subspaces, as the structure that represents the "elements of reality" in quantum theory. The quantum state is a derived entity, it is a device for the bookkeeping of probabilities (Pitowsky 2006; 214).

This quantum mechanical event structure imposes information-theoretic or probabilistic constraints on the set of events. From this viewpoint, QM is a theory about the probabilistic information obtainable from a set of measurements or gambles with a definite set of outcomes. The quantum state has an epistemic status in the sense that it represents our information or belief, encoded in probability assignments. It is a credence function that assigns probabilities to possible measurement outcomes. It should be noted that 'measurement' in this interpretation is a process in which a system *M* prepared in a specific way records information about some aspect of another system *S*. Therefore, in itself it does not imply subjective elements or any "high-level" characterization of observers (e.g. conscious ones). These elements, if anywhere, come into the assignment of probabilities, thus quantum states.

So, the Bub-Pitowsky interpretation is a non-representational information-theoretic interpretation of QM as a new theory of information or probabilistic correlations. QM should be understood as a theory which rests on the recognition of a new '*information-theoretic structure* to the mosaic of events' (Bub 2017). In other words, information has new structural features that allow for new ways of generating probabilistic correlations. From this viewpoint, QM can be regarded as a theory of relations or events which is in line with participatory realism. Furthermore, in this interpretation one accepts the perspectival nature of facts, as Bub's response to the Frauchiger-Renner argument testifies:

> On this option, what the Frauchiger-Renner argument shows is that *quantum mechanics, as it stands without embellishment, is self-contradictory if the quantum state is interpreted representationally*. The conclusion is avoided if we interpret the state probabilistically, with respect to a Boolean frame defined with respect to an "ultimate measuring instrument" or "ultimate observer." (Bub 2020; 254)

In the Frauchiger-Renner scenario (Frauchiger and Renner 2018), as formulated by Bub, there are two pairs of observers Alice-Bob and Wigner-Friend and it makes a factual difference which pair is regarded as the ultimate observer. Thus the necessity to make clear reference to such ultimate observer, when interpreting quantum states, forces us to deny that there are perspective-independent facts.

### 2.3. Janas-Cuffaro-Janssen Interpretation

Janas et al. (2022) put forward an interpretation of QM inspired by and similar to that of Bub-Pitowsky. First of all, they emphasize the non-ontic nature of the quantum state:

> In quantum theory, what is exhibited to us through the quantum state description *is not the set of observer-independent properties* of the system of interest. What is exhibited, rather, is the structure of, interrelations between, and interdependencies among *the possible perspectives that one can impose upon that system* (Janas et al. 2022; 232).

Moreover, the novelty of QM in contrast to classical mechanics stems from its novel kinematical structure:

> The kinematics of quantum mechanics presents us with a fundamentally *non-Boolean* probabilistic global structure. Upon this structure, in a given measurement context, we impose a particular Boolean frame. We do this to express our experience of the results of measurements conducted in that context (ibid., 214).

That is to say, there is no global Boolean frame into which every interaction with a quantum system could fit. In other words, QM does not describe a world of property-bearing objects interacting independently of us. Rather, it describes our connections or relations to a world that includes ourselves. Here, there is no room for detached observers. We are entangled with the world and the choice of a particular Boolean frame to represent a particular measurement context is up to us thus ruling out any perspective-independent description of the world.

### 2.4. Mueller's Interpretation

Mueller grounds his interpretation on the observation that different puzzles in the foundations of physics have a common core: they are all instances of the question "what will I see next?".[10] To answer this question one needs to start with the primitive notion of an 'observer state' which is a formalization of the information-theoretic state of the observer at any moment. Then, our theory can be summarized in the following formula

$$P(y \mid x) \qquad (1) .$$

That is, being in some observer state $x$ now, the theory should predict the probability of being in some other observer state $y$ next. According to this interpretation of QM,

> *all there is are observations* (i.e. transitions between observer states), and these observations are *non-deterministic*. Quantum states are the things that determine the probabilities of these observations (Mueller 2017; 26).

So, quantum states should be regarded as 'a catalogue of expectations about what an agent will see next.' Moreover, in this interpretation, there is nothing but transition events between observer states, following an interaction with an "observed" system. This is in line with a participatory realist metaphysics of events as interactions. And finally this interpretive approach takes the first-person perspective of observers as fundamental thereby acknowledging the perspectival nature of facts, and allowing for their relative nature. Here, the notion of observer can in principle refer to any physical system. Furthermore, an observer is not characterized as an individual object but in terms of its observer states. Also this rejection of object-based ontology is in keeping with participatory realism.

It should be noted that Mueller's work is a toy model and not to be regarded as a full-fledged interpretation of QM. Moreover, his work on the foundations of QM is

[10] . It should be noted that Mueller's work is a toy model and not to be regarded as a full-fledged interpretation of QM. Moreover, his work on the foundations of QM is situated within a more general discussion of the relation between computation and physics.

situated within a more general discussion of the relation between computation and physics.

### 2.5. Relational Quantum Mechanics

In RQM, the quantum state ψ encodes the information that an observing system $A$ has about an observed system $S$ (Laudisa and Rovelli 2021). This information is acquired via past interactions between these two systems and so the quantum state ψ "of S" is meaningful only relative to $A$. Moreover, the quantum state ψ is a tool that can be used by $A$ to predict the outcomes of the future interactions between $A$ and $S$. Thus, in the Heisenberg picture, the dynamics of QM describes

> discrete changes of the relative state, when information is updated, and nothing else (Smerlak and Rovelli 2008; 3).

Therefore, in this interpretation, QM is the theory of 'logical relations' between the epistemic and predictive aspects of physical information. The epistemic aspect relates to the relative nature of the quantum state ψ and its predictive aspect concerns the role of ψ as a tool for prediction of future outcomes of interactions. So, laws of QM are not describing the absolute, intrinsic properties of systems and their evolution through time. Rather,

> physics concerns what can be said about nature on the basis of information that *any* physical system can, in principle, have (ibid., 2).

So, quantum states and values of physical quantities associated to a system only make sense in relation to another physical system, called the observer or reference system. Moreover, physical variables take values only in interactions between two different systems. Note, however, that there is no reference to conscious agents or observers: in RQM the second "observing" system could be strictly any other physical system.

Since, according to RQM, physical variables take values only in interactions between two systems,

> a good name for the actualization of the value of a variable in an interaction is 'quantum event'. The proper ontology for QM is a sparse ontology of (relational) quantum events happening at interactions between physical systems (Rovelli 2018; 7).

That is a participatory realist viewpoint according to which QM is primarily about events associated to interactions; consequently the concept of a physical system is secondary. A physical system is defined as a bundle of relations, lacking any intrinsic properties. This is consistent, but leads to the question of what should we say about a non-interacting system? One option is to accept that the notion of the state of an isolated, non-interacting system is simply meaningless, since no property can be attributed to it, missing the interactions that alone could produce them. The only alternative seems to be, as Dorato (2016) argues, that such a system can be said to only possess intrinsic dispositional properties to manifest definite values of the corresponding quantities in interaction with other systems.

Another crucial element of RQM is that all physical systems are regarded as equivalent without any fundamental distinction between an observing system and an observed system (Rovelli 1996). Any physical system whatsoever can be regarded as an observing system *relative* to any other observed system. Thus, from this viewpoint, QM is a theory describing the general form of the information that any system can have about another system (van Fraassen 2010). Note the thoroughgoing objective relationalism of RQM. It is an interpretation of QM in terms of the objective relations between physical systems and the correlations/informations generated upon their interactions. It aims to avoid any kind of reference to conscious agents and their epistemic states altogether or, in fact, to agency tout court, unless this is entirely objectified such that any physical system, no matter how elementary, may count as an "agent" (or observer).

In RQM, the quantum state and the properties of a system are only meaningful relative to another system. Therefore,

> quantum mechanics indicates that the notion of a "universal" description of the state of the world, shared by all observers, is a

> concept which is physically untenable, on experimental grounds (Rovelli 1996, 1650).

That is, there is no universal or perspective-independent account of events or situations. Facts make sense only within a perspective. When framing Wigner's friend scenario in RQM, 'Wigner's facts are not necessarily his Friend's facts' (Di Biagio and Rovelli 2021). However, as Adlam and Rovelli (2022) argue, observers can reach intersubjective agreement via cross-perspectival links. This is in keeping with perspectival objectivity. According to the postulate of cross-perspective links, the information stored in Alice's physical variables about the variable V of the system S upon their interaction is accessible in principle to any observer who measures her in the right basis, so this information can be regarded as an invariant element of RQM. It should be emphasized that the existence of an isomorphism between different perspectives is sufficient for perspectival objectivity to hold. This guarantees intersubjective agreement. The existence of invariant elements, though not a necessary requirement for perspectival objectivity, can be interpreted as observer-independent facts about reality. This is, of course, not in tension with perspectival objectivity and can be regarded as an additional feature of this framework resulting in a stronger form of objectivity.

### 2.6. Brukner-Zeilinger Interpretation (BZI)

In the BZI, the epistemic (and observer-dependent) character of the quantum state is explicitly acknowledged:

> All the "quantum state" is meant to be is a representation of that catalog of our knowledge of the system that is necessary to arrive at the set of, in general probabilistic, predictions for all possible future observations of the system (Brukner and Zeilinger 2002; 8).

Here it should be noted that the quantum state is defined with respect to a real or *hypothetical* observer. In other words, one can apply QM in counterfactual situations. For example, in assigning a quantum state to a situation in the early universe we are saying that a hypothetical observer in that situation would assign that particular state given the available information. Moreover, the assignment of a quantum state requires a movable but necessary epistemic cut between the observer and the observed system. The universality of QM implies that any observing system and its corresponding observed system can be treated as an object of measurement by a second observer (thus a third system). So, in a given measurement process, the observer is not modeled by QM. This makes the measurement process irreversible *for all practical purposes*, from an epistemic perspective. However, the physical interaction between the observer and the measured system is *in principle* reversible simply because QM is a reversible theory (Brukner 2017).

The second feature of BZI is a participatory realist ontology. According to Zeilinger (1999) our physical description of the world is represented by propositions. We describe a physical system by a set of true propositions. Now, an important question is: how do we obtain the propositions describing a system or object and how do we verify them? His answer is that we obtain such propositions through earlier observations and verify them through future observations. Thus, all the properties (expressed in propositions) which we assign to a system or object are obtained through observations and are credible only insofar as they do not contradict further observations. An observation or measurement is an interaction between the experimentalist through the experimental arrangement with the physical system. According to Brukner and Zeilinger (2002) 'the experience of the ultimate experimenter is a stream of ("yes" or "no") answers to the questions posed to Nature'. In line with participatory realism, these discrete observational events lie at the foundation of the ontology of BZI and everything else is constructed out of them. Therefore, the physical system or object is 'a useful construct connecting observations' (Zeilinger 1999; 633), i.e. it comes into existence as a result of such observations, not being coherently characterized outside of them.

The third feature of BZI is the perspectival nature of facts as discussed in the context of the Wigner's friend paradox. Given that both Wigner and Wigner's friend can experimentally verify their respective descriptions, it seems plausible to regard them as observer-independent facts or 'facts of the world'. However, one can prove interesting no-go theorems for observer-independent facts, showing that one cannot define joint probabilities for Wigner's outcome and for that of Wigner's friend. A possible interpretation of this is that

> there cannot be facts of the world per se, but only relative to an observer (Brukner 2018, 7).

This is the perspectival or relative nature of facts shared by all the pragmatist epistemic interpretations of QM.

### 2.7. QBism

QBists advocate an epistemic view of the quantum state and an interpretation of this epistemic nature of a subjectivist kind. They maintain that quantum states are degrees of belief of an agent concerning the outcomes of her interactions with the world. A quantum state is a catalogue of beliefs.

In QBism, events (or "facts") are elements of reality with respect to which two poles are defined: the agent and the experienced object or system, not possessing an independent, individual characterization outside such events (Pienaar 2020). However, the occurrence of events is a ubiquitous phenomenon not necessarily requiring an agent-object relationship:

> I think the greatest lesson quantum theory holds for us is that when two pieces of the world come together, they give birth. [Bring two fists together and then open them to imply an explosion.] They give birth to FACTS in a way not so unlike the romantic notion of parenthood: that a child is more than the sum total of her parents, an entity unto herself with untold potential for reshaping the world (Fuchs 2017a, 122).

That is, events are 'moments of creation' upon the interactions between physical systems. The interaction between a decision-making agent and a physical system, i.e. a measurement event, is just a special case of the occurrence of an event. This ontology of events or relations is the participatory realist element of QBism. It is worth emphasizing that QBists are not instrumentalist or anti-realist. 'Rather than relinquishing the idea of reality (as they are often accused of), they are saying that reality is *more* than any third-person perspective can capture. Thus, far from instances of instrumentalism or antirealism, these views of quantum theory should be regarded as attempts to make a deep statement about the nature of reality' (ibid., 113).

Also the perspectival nature of facts occupies a central place in QBism. For QBists, Wigner's friend paradox shows that we should recognize as fundamental rather than paradoxical the idea that outcomes of experiments are necessarily personal experiences of the agent carrying out the experiment, rather than "objective" or "observer-independent" facts. In Wigner's friend scenario, Wigner and his friend assign different quantum states to the same quantum system. Now the question is this: which one is correct? If the outcome of an experiment is an invariant fact available for anyone's inspection, why should Wigner and his friend assign different states to the same system? One might answer that they have different states of information. But, who has the right state of information? According to QBism, the whole puzzle stems from insisting on taking what are personal experiences as agent-independent objective facts. The same applies to similar apparently paradoxical situations, like in the Frauchiger-Renner scenario. For QBists, facts are experiences of agents interacting with the world. However, QBists do not deny the possibility of communication and so the construction of a public world. 'In this way a common body of reality can be constructed, limited only by the inability of language to represent the full flavor—the "qualia"—of personal experience' (Fuchs et al. 2014; 750). This suffices for perspectival objectivity to hold.

### 2.8. Healey's Pragmatist Interpretation

According to Healey, quantum states do not describe or represent the properties or relations of any physical system. Rather,

> the constitutive function of quantum states is to prescribe objectively rational degrees of belief concerning events to (actual or hypothetical) physically situated agents (Healey 2017; 71).

So, quantum states through the Born rule give us probabilities for the occurrence of events. The quantum probabilities are about *non-quantum magnitude claims* (NQMCs) which are sentences of the form S: 'The value of observable *A* on system *s* lies in the range $\Delta$' (Healey 2012). The predictions and explanations of QM are expressed in the form of NQMCs. So, QM advises agents assigning quantum states which NQMCs are licensed to make and which are not. In other words, QM advises the agent what degree of belief or credence to attach to NQMCs.

This pragmatist approach seems compatible with a participatory realist viewpoint. QM advises the agent on what to expect and how to act in different physical situations. And this interaction between the agent and the world is an event expressed in the form of NQMCs. Moreover, as Healey emphasizes, QM itself has no beables. A 'beable' of a theory is what exists in three-dimensional space and could be described in classical terms (Bell 2004). Instead, QM has assumables, i.e., 'things whose physical existence is presupposed' by an application of a quantum model (Healey 2017; 127). A measurement having a definite outcome and the readings of the instruments registering the outcomes of measurements are among the important assumables of QM. Clearly, this is a move away from object-based ontology and towards a relation/event ontology in line with participatory realism.

Also in Healey's pragmatist approach a perspectival nature of facts is acknowledged. Referring to the Wigner's friend paradox and its extensions:

> there is an ideal of objectivity that is not met here, according to which a claim correctly ascribes an objective property to something just in case the world contains a corresponding object which intrinsically possesses that property when represented as it is in itself, irrespective of its environment and not just from the perspective of physically situated and cognitively limited agents (ibid., 199).

Moreover, Healey relies on environmental decoherence and Quantum Darwinism to show how differently situated agents all *agree* on the content of the magnitude claims about a system, thereby securing their objectivity. So, QM puts limits on the extent and nature of objectivity. It questions the existence of absolute observer-independent facts (Healey 2018). As discussed above, on this account, the only sort of objectivity we can achieve is intersubjective agreement or covariance between different perspectives.

Lastly, it should be emphasized that regarding the observing system, Healey is walking on a tightrope between QBism and RQM. All of them share the relational view of quantum states. For QBists the quantum state depends on the epistemic state of the agent assigning it while for Healey it depends on the physical circumstances defining the perspective of the agent. However, if Healey stops here then his position will not differ from that of RQM. According to RQM's postulate of cross-perspective links, the information stored in Alice's physical variables about the variable V of the system S upon their interaction is accessible in principle to any observer who measures her in the right basis, so this information can be regarded as an invariant element of RQM (Adlam and Rovelli 2022). Similarly, for Healey (2012), the content of a NQMC about a system *s* does not depend on the agent situation and so can be regarded as offering an objective agent-independent physical description. The only difference between Healey and RQM is that, for Healey, the acceptance of QM limits the contents of some NQMCs in such a way that they no longer contribute to the explanatory and predictive goals of physics and so are better left unsaid.

Moreover, every physical situation can define a perspective and so it is not clear what is meant by 'hypothetical agency'. If 'hypothetical agency' is just a fancy name for the perspective of a physical system or situation then it is just the observing system of RQM. If it is something more, then quantum states and probabilities will certainly depend on agents whether human, merely conscious, or neither. If quantum states and probabilities could exist without agency then Healey's position, in spite of his pronouncements to the contrary, collapses to that of RQM. Unfortunately, Healey is not clear on this point. Moreover, though quantum states and probabilities for Healey are objective and relational, they are not ontic. The amount of money I owe is not subjective. However, it does not exist without agents like me. Objectivity and onticity are not the same.

## 3. Interpretation underdetermination

So far we have argued that there are several interpretations of the quantum mechanical formalism that may all be analyzed in terms of the degree to which they emphasize (and in some cases are based on) the three theses of an epistemic view of the quantum state, a metaphysics of participatory realism, and an epistemology of perspectival objectivity. We have done so by directly pointing to the original articulation of such interpretations, where the above elements are stated, albeit in different forms. It is worth emphasizing that our purpose is not to simply review and explain what the proponents of different approaches have written and what they mean, as if reviewing the (recent) history of science. And because of this, our points would stand also if some of the authors would disagree on the idea of fitting in a common framework, because they would rather emphasize their differences with other approaches. Our purpose is to suggest that these views *can* be put under a common "interpretative framework", shown to share some key aspects with deep philosophical significance, and then also analyze where the differences between them can be found (including the possibility that, in some cases, these differences will be found too significant to keep them inside the same framework). In this sense, we are charting the extent to which one has true underdetermination as well as identifying the novel directions for philosophical investigation they indicate.

Indeed, despite this common core of interpretive commitments there are important points of difference among these epistemic-pragmatist interpretations. As we will discuss below, some of them can be considered as "differences in degree" or in emphasis about one or the other commitment, or about their implications. Others are more substantial, and correspond to more fundamental disagreements about how the same general commitment should be understood or which implications should be drawn from it. These disagreements cannot be resolved as just different ways of expressing the common core.

One can classify these more profound differences between different pragmatist epistemic interpretations in the following way: a) differences that can be formulated and in principle resolved within the formalism of QM and; b) those that require one to go beyond the formalism either to resolve them or to recognize them as foundational, leading to a further classification of epistemic-pragmatist interpretations, based on a further articulation of the common core.

The nature of the observing system and the epistemic nature of the quantum state belong to the first class. Should we be allowed to regard any physical system as a possible "observing" system, as it enters the common core, or do we need to include a more refined notion of agency in its definition? Similarly, is the quantum state a state of information or a state of belief? It is imaginable that some theoretical result in the form of a no-go theorem can decide between these alternatives; examples of such no-go theorems can and will be provided.

On the other hand, the interpretation of probabilities and the concept of a law of nature belong to the second class. What are the conditions for a system to be able to use the laws of QM? and what are physical laws in general? and what notion of probabilities is compatible with the other tenets of a given QM interpretation? Attempts to answer questions like these force us to go beyond purely quantum mechanical considerations, into the domain of more general philosophical debates.

As a last point, it is clear from our exposition of the epistemic-pragmatist approaches that they differ in the precise understanding of the measurement outcomes. However, all of them regard the measurement outcomes as relational and the results of interactions between the observing system and the observed system. These differences can be found to be different articulations in detail of this (and other) shared general points of philosophical significance even before the further articulation. A similar situation (shared general aspects, different articulations, leading to different takes on measurement outcomes) has been suggested for other approaches (in the ontic camp) by Callender. He regards the contrast between Bohmian mechanics, GRW and Everett as a case of genuine underdetermination in QM despite no common understanding of the measurement outcomes. As he puts it, 'nothing compares classically, for example, to the contrast between the sparse ontology of GRW (with 'flash' ontology) and the generous ontology of Everett' (Callender 2020, 58). Thus, measurement outcomes are 'flashes' in fGRW; 'masses' in mGRW; 'positions' in Bohmian mechanics and 'agent's experiences' in the subjectivist approach to probability in MWI. So, we don't think there has to be a common base of quantum measurement outcomes for underdetermination to obtain. If one were to insist on this as a requirement for underdetermination in QM then there would be no such thing in QM as there is no agreement on the interpretation of the measurement outcomes that are supposed to provide the evidence for QM.

In this section, we focus on what we feel are the most important differences between the pragmatist-epistemic interpretations, namely the following interrelated issues:

1. The nature of the observing system
2. The interpretation of probability

As we argued earlier, according to the epistemic interpretations, QM is about the interactions between an observing system and an observed system. The quantum state is defined only relative to the observing system. Thus, one important interpretive task is to clarify the nature of the observing system. Moreover, the quantum mechanical formalism gives probabilistic predictions for the results of the interactions between the two systems. In fact, one can express the whole quantum mechanical formalism purely in terms of probabilities and without ever invoking quantum states or Hilbert spaces. This implies that one's interpretation of probability is closely related to one's interpretation of the quantum state and vice versa. One should give a consistent answer to both interpretive questions, unless one decides to argue that only a specific formulation of the quantum formalism is meaningful, based on purely philosophical considerations and despite any mathematical equivalence.

### 3.1. The nature of the observing system

Let us begin our discussion with the nature of the observing system that the quantum state refers to, in different epistemic-pragmatist interpretations.

In the Bub-Pitowsky style interpretation, to begin with, the experimental set-up plays the role of the observing system (Pitowsky 2006, 236). There is no reference to the experimentalist herself as an agent or any special restrictions on the observing system as a quantum system. It is replaced with or reduced to a part of the context of experiment. And everything is referred to the possible results of interactions between the observing system and the observed system as parts of the experimental set-up. One would then conclude that some high degree of complexity and possibly a classical nature would have to be included in the characterization of the observer (Bub 2020, 254). This is also largely in line with Bohr's view.

In RQM, it is emphasized that the relational viewpoint on QM can be interpreted solely in terms of the relationship between the observing system and the observed system *such that* the observing system can be *any physical system* with any degree of complexity. Even a physical system as simple as an electron can be considered as an observing system in the sense that its past history is in principle sufficient for computing the outcomes of its future interactions with other systems and this is what is encoded in the quantum state (Smerlak and Rovelli, 2008). So, we have a radically relational but also strongly "objectified" and "naturalized" interpretation of QM. No "subject", no distinctive "agency" (beside the possibility to interact with other physical systems), is called for. There is nothing in this picture except interactions between different systems all equivalent to each other in their status of possible observers and in the fact that they can participate in the assignment of a quantum state.

The status of the observing system in Mueller's interpretation is similar to that of RQM. In this interpretation, the observer is a general notion characterized by its information-theoretic state. Actually, for Mueller (2017, 24), an observer 'is its observer state.' In other words, an observer can be any system described by a certain information-theoretic state at a given moment.

In contrast to this, according to Healey's pragmatist approach,

> a quantum state ascription is not relative to an arbitrary distinct quantum system, but rather to the perspective of an actual or potential *agent*—some physically situated *user* of quantum theory (Healey 2012; 22).

Of course, an agent or a physically situated user of QM can be regarded as a quantum system but not every quantum system is an agent or a physically situated user of QM.

Similarly, in BZI, the observing system is a real or *hypothetical* observer with definite experimental capabilities. The notion of hypothetical observer or potential agent signifies the applicability of QM in counterfactual situations. That is, if there were such and such a situation then any observer in that situation would assign the same quantum state or have the same degrees of belief. Moreover, according to Brukner (2021)'s no-go theorem, not only the possibility to be an "observer" of another physical system, and thus participating in the assignment of a QM state, is not always granted to any physical system, but it depends on *relative* features of the two systems under consideration. More precisely, the observing system should be able to distinguish between the alternative states of the observed system. Thus, a qubit can act as the observing system for another qubit because it has the resources for distinguishing the two states of the other qubit. But, a qubit cannot act as an observing system for a quantum system with more than two degrees of freedom because it does not have the necessary resources for distinguishing more than two degrees of freedom. So, the observing system should have at least enough degrees of freedom to distinguish between the degrees of freedom of the observed system, since it is this set of degrees of freedom that is encoded in the quantum states assigned to the observed system (in relation to the observing system). The observing-observed relationship between quantum systems is a relative notion. We note that this constraint on the required resources of the observing system could also be articulated in terms of (relative) complexity. Doing so is an important point deserving further analysis, which we leave for the future.

We also note that, just like in Healey's perspective, the "observer" system can be entirely naturalized, partially or entirely embodied in a measurement apparatus or context, and reduced to its physical aspects (no "subjectivity" or "internal experience", let alone "consciousness"); simply, it cannot be *any* physical system.

Pursuing this line of thought, in the same direction as both Healey's pragmatism and BZI, QBists add more constraints on the observing system. Not only it cannot just be any physical system, but it should be able to use the QM formalism *normatively*. And for this, it should enjoy some higher level of agency in the sense that it can act (freely) on other systems in the world so that the consequences of its actions matter for it (DeBrota et al., 2020). More importantly, it should be capable of having experiences. This is so because for QBists the measurement outcome is first of all an experience for the agent doing the experiment (Pienaar 2021). From this perspective, QM is precisely a tool for such decision-making agents that they can use to make better decisions in their (future) interactions with other systems.

Therefore, to summarize, within the broader epistemic-pragmatist class whose characterizing elements we tried to identify, there is an interpretive spectrum regarding the nature of the observing system with QBism and RQM at its opposite ends. Both QBism and RQM interpret quantum theory in a relational manner as the relationship between two systems. In RQM, it is the relation between any two physical systems. In QBism, this is the relation between an (high-level) agent and the observed system. One important point to add is that in all approaches along this spectrum, the "observing" system can be to a large extent "objectified" or "naturalized", even when it corresponds to "agents" in a lab, in particular in terms of a measurement apparatus (or other material entity) used in its interaction with the "observed" system. This is maybe obvious for the other interpretations, but, importantly, even for QBists a measuring instrument is literally an extension of the agent (Fuchs 2017b). As soon as one objectifies the agent by regarding the apparatus and the experimental context as extensions of the agent, the difference between the QM interpretations amounts to different specifications of what can *not* be embodied in the definition of agency, or which minimal complexity should the embodied elements have, or how constrained by physical situation are the actions of the agent. This is where it is in principle possible for theoretical results in the form of no-go theorems to favor one of the interpretations, by putting constraints on the embodied aspects of the observer. It is the first way in which the choice of QM interpretation within this class, which, we recall, is based on the assumption that the QM formalism should not be modified, can be affected by developments within the theory itself.

### 3.2. The interpretation of probability

Let us turn to the interpretative issues surrounding quantum probabilities, which help differentiating between QM interpretations of the epistemic-pragmatist class and, we argue, represent another philosophical basis on which the choice between them can be made.

It is known that one can mathematically express the quantum state purely in terms of conditional probabilities for the results of all possible measurements performed on the system it is assigned to, and that QM as a whole can thus be recast in a dynamical theory of such probabilities. From this perspective, it is clear that one's view on the nature of probabilities should at the very least be compatible with one's view of quantum states and QM more generally, if not entirely determine it. Moreover, as we will see in the following, the discussion on the nature of probabilities as applied in an observer-observed context is intertwined with that of the nature of the observer system, which we have seen in the previous section. The observer system is one that could apply probabilities and so, in a sense, it is the nature of probabilities that defines the nature of the observer system. An observer system that can apply Bayesian probabilities is different from one that can only use frequencies.

There are basically two perspectives on the nature of probabilities that are general enough to apply to all cases where QM is applied in theoretical physics, which include single-instance events and where idealizations about experiments (like those that are necessary for a frequentist approach) are better avoided. One is the subjectivist Bayesian view, in its various (more or less "objectified") versions; the other is the view in terms of propensities. Before discussing how probabilities are viewed in different epistemic-pragmatist QM interpretations, let us recall briefly some elements of the above views on probabilities.

According to Bayesianism, probabilities are measures of the subjective degrees of belief about future events of agents. However, they also encode the knowledge or evidence possessed by the agent holding those beliefs, which are thus somehow constrained by them (see e.g., Myrvold 2021). At the opposite end, in the propensity view, probabilities are objective properties of physical systems indicating their disposition to possess a given feature or evolve in a certain way. But things are more subtle, both from the Bayesian and from the propensity perspective.

There is a spectrum of interpretations of probabilities from the fully subjective Bayesian view, across more objective versions of the same, toward the fully objective propensity interpretation. If one takes the fully subjective Bayesianism at one end of this spectrum and add to it all the constraints originating from rationality[11], evidence and communication then one arrives at the objective Bayesianism which comes very close to the other end of the spectrum, i.e., the propensity interpretation. Let us analyze this transition from a purely subjective Bayesian interpretation of probability to an objective propensity interpretation of probability in more detail. According to the subjective Bayesian view, a probability assignment is just a personal judgment:

> Where personalist Bayesianism breaks away the most from other developments of probability theory is that it says there are no *external* criteria for declaring an isolated probability assignment right or wrong. The only basis for a judgment of adequacy comes from the *inside*, from the greater mesh of beliefs [the agent] may have the time or energy to access when appraising coherence (Fuchs and Stacey 2019; 6).

[11] . By ‘rationality’ here we mean the set of rules and practices (similar to Kuhn’s paradigm or disciplinary matrix) accepted and followed by the scientific community in a specific period making consensus possible.

Degrees of belief can differ from agent to agent. The only constraint on the agent's degrees of beliefs is that the whole mesh of her beliefs should satisfy a coherence condition. The agent should strive to assign her probabilities so that they will not lead to a sure loss. To be coherent, the agent's personal probabilities should obey the rules of probability. This is a rationality constraint. Let us call this the first-level constraint. However, there is a second-level constraint relating to the available evidence. We can admit that probabilities are degrees of belief but deny their purely subjective or personal character. We can argue that given the available evidence regarding a situation there is an objective probability to be assigned by any reasonable agent. This is the objective Bayesian interpretation of probability. From this viewpoint, probabilities are objective in the sense that they depend on the available evidence. They are about 'what it is reasonable for any reasonable person to believe' given the available evidence (Hacking 2009). Furthermore, we can identify a third-level constraint relating to communication and agreement within a community, which concur in the determination of the "right" probabilities to assign in any given situation.

Whatever the specific version of Bayesianism one adopts, it is clear that the Bayesian view requires agency of some sort and that it is not compatible with a definition of the agent that does not endow it with some minimal complexity, since it relies on notions like knowledge, inference or belief. Thus, it is also clearly a relational view: Bayesian probabilities cannot be thought of as intrinsic properties of the system they refer to.

The "probabilities from propensities" view is instead one in which probabilities are objective and do not require agency for their definition. Also in this case, however, the use of propensities to characterize (probabilistically) the "observed system" has consequences for the "observing system". Let us expand.

Following Gillies (2000) we can classify the propensity interpretation into the *long-run propensity theories* and the *single-case propensity theories*. According to the long-run propensity theory, propensities are associated with *repeatable conditions* which in a long series of their repetitions are disposed to produce frequencies approximately equal to the probabilities. On the other hand, in a single-case propensity theory, propensities are regarded as dispositions to produce a particular outcome on a specific occasion. For Popper (1995), these propensities are properties of the whole physical situation which can ultimately include the whole state of the universe at that particular occasion. Also for Miller (1994) propensities are dispositions of a state of affairs to realize an outcome. Therefore, ‘propensities are not located in physical things, and not in local situations either. Strictly, every propensity (absolute or conditional) must be referred to the complete situation of the universe (or the light-cone) at that time’ (ibid.). Fetzer (1982) weakens this requirement and regards propensities as depending on a complete set of *relevant* conditions on a specific occasion. Therefore, according to the propensity interpretation, propensities are not regarded as the properties of an individual physical system or object but always refer to a context which includes, is specified by or partially defines the observer system. This also implies that the assignment of probabilities requires a level of complexity that is larger than that given by the observed system, and part of the definition of an observing system. By changing the context, a change in the observing system, or a change operated by the observing system, changes the probabilities for the results of its interactions with the observed system. This context-dependence and relational nature of probabilities, and the consequent role and complexity of the observing system, is crucial for epistemic-pragmatist approaches to QM, and it has important implications for the relationship between the Bayesian and the propensity interpretations of quantum probabilities.

Let us now start examining the interpretation of quantum probabilities in epistemic-pragmatist approaches to QM. We do so by first contrasting the 'polar opposites' in this camp, RQM and QBism, starting from RQM.

In the RQM literature, e.g. in the analysis of the EPR correlations (Smerlak and Rovelli, 2008) one finds calls for a subjective interpretation of probabilities, and this is often accompanied by the use of notions like “knowledge” and “inference”. This would suggest a richer kind of observers/agents (if not conscious ones). However, RQM is introduced as an objective relational interpretation of QM only concerned with the information that generic physical systems (including, e.g. two electrons or two qubits) can have about each other.  Indeed, in the same literature one also finds explicit statements cautioning against "reading too much" in the use of terms like knowledge, epistemic or observer in the RQM context (Di Biagio and Rovelli, 2021b), since this would contradict the basic tenets of RQM concerning the quantum state and the "observing system" it refers to.

That is also why Dorato (2016) suggests that, in order to escape inconsistency, RQM should avoid any talks of subjective probability which by definition refers to epistemic states of agents. Instead, subjective probabilities should be replaced with 'objective, mind-independent dispositional probabilities (propensities).' This avoids blatant contradictions, but still raises concerns.

If probabilities in RQM are interpreted in an objective manner as propensities or dispositions then it means that the quantum state cannot be regarded as epistemic in RQM. Indeed, probabilities as dispositions are by definition part of the ontic structure of the world, and the quantum state can be seen as equivalent to a list of such probabilities. Therefore, the quantum state should be regarded as part of the ontic structure of the world too. This deduction would bring RQM, in fact, outside the epistemic-pragmatist camp as we characterized it, and would also runs contrary to several claims in the RQM literature itself explicitly denying any such ontic nature for quantum states, stressing that they represent nothing more than "tools" to compute probabilities of future events (Laudisa and Rovelli, 2021). In the end, it seems that we run into some fundamental difficulty, calling for more work to achieve a consistent picture compatible with all the RQM desiderata.

In passing, we also notice that an ontic view of quantum states would bring RQM closer to the MWI of QM, perhaps not surprisingly, given the common Everettian origin of both. Interestingly enough, though, while starting from an ontic view on quantum states, some versions of the MWI come close to the pragmatist epistemic interpretations in terms of their interpretation of probabilities. Some of them, in fact, adopt a subjective interpretation of probabilities, being measures of the ignorance (or knowledge/belief) of the observer with regard to its location at a specific branch of the quantum state (Vaidman 2012; Saunders 1998). Although, it should be noted that, despite this earlier take about probabilities in the context of MWI, more recently, Saunders (2024) defends a form of finite frequentism as an interpretation of quantum probabilities.

Let's move to the other end of the pragmatist-epistemic spectrum. In contrast to a propensity interpretation of probability in RQM, QBists explicitly endorse the subjective or personalist view of probabilities: they are personal Bayesian degrees of belief which express the agent's expectations for her future experiences. Probabilities are assigned to an event by an agent and can be defined in terms of her betting behavior. This interpretation is certainly consistent, in its combined views of observing system and probabilities. Its radical subjective character has the benefit of simplicity, but is also at risk of making it less natural to ask further questions about eventual constraints that the agent assigning probabilities has to satisfy, or how her assignment depends on past history, evidence or physical limitations. These become, however, the interesting issues in further articulating a QM interpretation, once a Bayesian view on probabilities, and its implications for the quantum observer, is adopted. Other QM interpretations in the epistemic-pragmatist class try to take on board such constraints.

The interpretive status of probabilities in BZI is a bit ambiguous. There are passages that indicate a frequency interpretation of probability in BZI:

> In a measurement, one of the possible measurement results becomes reality with a relative frequency as indicated by the quantum state (Zeilinger 2002, 252).

Similarly, Brukner and Zeilinger (2000) state that in an experimental arrangement with $n$ possible outcomes and probabilities $p_j$ for the outcomes '*all* an experimenter can do is to guess how many times a specific outcome will occur.' However, they maintain that due to the statistical fluctuations in any finite number of repetitions of the experiment this relative frequency is not precisely predictable. Instead, the experimenter's degree of uncertainty or lack of information in the relative frequency of a certain outcome is computable. The same uncertainty is inevitable of course also when considering single observations or experiments, which is also where one has to go beyond frequentism in the interpretation of probabilities. Consequently, they derive the following formula for 'the lack of information about the outcome $j$ with respect to a *single* future experimental trial':

$$U\left(p_j\right) = p_j(1 - p_j) \quad (2)$$

This reference to 'the lack of information' suggests a subjective interpretation of probability. Thus, even restricting to lab-based situations, the experimenter can verify her probability assignments by measuring relative frequencies, but probabilities themselves are not relative frequencies. Rather, they measure the experimenter's degree of uncertainty or lack of information about the outcomes of the experiment. Thus, there is a subjective or belief-type aspect in the interpretation of probability in BZI, approaching QBism. However, it differs from QBism in the fact that they interpret probabilities in an objective Bayesian sense, in which the available evidence from the experimental context fixes (or at least strongly constrains) the probabilities to be assigned, and they are moreover further constrained by rationality conditions (since BZI only consider explicitly how QM is used by observers in a (idealized) lab-like situation).

The full set of constraints that constrain, if not determine, probability assignments in a Bayesian context is taken into account and discussed in detail by Healey in his pragmatist QM interpretation (Healey 2012; Healey 2018).

In Healey's view, in fact, rationality constraints, those coming from evidence and even those coming from the "social" agreement within the community of scientists come to determine the (still Bayesian) assignment of quantum probabilities, that is, of quantum states to observed systems. Quantum probabilities are therefore objective in the sense that,

> Quantum theory itself commands that quantum probabilities conform to the Born rule: the community's collective evidence-based judgment commands use of a particular quantum state in the Born rule. Neither command is arbitrary: the authority in each case rests ultimately on experimental and observational results and collective judgment of their evidential bearing (Healey 2012, 9).

It should be emphasized that this objectivity does not modify the relational nature of quantum states; for Healey (2017) both the quantum state assignments and the probabilistic statements should be understood as relativized to the physical situation of an actual or hypothetical agent. So, on the one hand quantum probabilities are objective in the sense that there is a unique probability to be assigned in a physical agent-situation. On the other hand, they are relational in the sense that they are meaningful only relative to a specific physical agent-situation.

Thus, the spectrum of views concerning probabilities in general reflects a similar spectrum of views on quantum probabilities, that we can recognize in the epistemic-pragmatist camp. Now, let us further compare the objective Bayesian view of probability with the propensity interpretation of probability. The former emphasizes the process of construction by the agent to arrive at a probability judgment which includes mental, sociological and historical elements and the latter emphasizes the final product of this process of construction as an objective fact of the world. According to the propensity interpretation, propensities are context-dependent and therefore relational. They are objective in the sense that a given context determines a unique propensity. But this notion of 'context' is actually the product of a long process of construction on the part of the agent. It includes the mental construction of the concept of a system as an object isolated from the rest of the world and also the physical construction of an experimental arrangement as an action on the world. In the propensity interpretation, one disregards this whole process of construction and emphasizes the final product as an objective fact of the world. But, even this notion of objectivity itself is the product of a process of construction at the individual and social level. Objectivity has social aspects. Self-criticism by the individual and debates and criticisms among the individuals in the scientific community through a long historical process are instrumental in constructing what we regard as objective. This requires us as a community to *agree* to follow certain rules and respect certain norms. Thus, whenever we talk about objectivity in the present context, the notions of normativity and convention are always present.

So, at the level of probabilities, one can start with the purely subjective Bayesianism of QBism and by objectifying it one will arrive at objective Bayesianism which is very close to the propensity interpretation that could be associated with RQM, although it still requires a more refined, more complex notion of agent-observer system (how much, though, is still to be determined in more precise terms). Actually, one can appeal to Lewis (1986)'s Principal Principle to understand the transition from a subjectivist to an objectivist interpretation of probability within the formalism of QM. The PP connects an agent's degree of belief in event *e* at time *t* to the objective probability of *e* at *t*. Taking into account all of the evidence for an event, a rational agent's degree of belief in the event should be equal to the corresponding objective probability for that event. However, it should be noted that the PP as a constraint on the agent's belief can do its job only under the assumption that objective probabilities exist, which is of course far from granted. Moreover, as Healey (2017) points out, by itself it does not address the question of how an agent can know or have evidence for the chance of *e*.

Thus, we see that the interpretive differences concerning probability can be investigated at different levels of analysis even though there is nothing within the formalism of QM to favor one over the other. Firstly, we see the difference in emphasis on the role of the evidence and other constraints on the degree of belief of the agent in Bayesian interpretations of probability. However, beliefs essentially remain different from the propensities or dispositions in the objective interpretation of probability. As we noticed, this is a substantial difference that affects the status of probability and, by implication, of the quantum state in RQM with respect to other QM interpretations in the epistemic-pragmatist class. Secondly, the undecidability of these different approaches within the formalism of QM by no means reduces their conceptual significance. Every epistemic-pragmatist approach comes equipped with its own interpretation of the epistemological and metaphysical elements of this framework resulting in different conceptual and philosophical issues worth analyzing and comparing in their own right. This conceptual analysis presupposes the formalism of QM and explores the philosophy of QM from the perspective of epistemic-pragmatist approaches. Thirdly, it is in principle possible for there to be theoretical results in the form of no-go theorems that compel us to favor one interpretation over the other, though not ruling out any interpretation altogether, by constraining the kind of physical system that can play the role of "observer" in the relational determination of quantum states or, related to this, of probability-assigning agent. For example, there is Brukner (2021)'s no-go theorem that Qubits are not observers. If successful it poses a challenge for RQM according to which the observing system could be any system whatsoever.

As a last point, it is worth mentioning a possible objection to our line of argument. We have taken for granted that quantum states are conceptually or physically equivalent to a set of probabilities because mathematically they can be shown to be so. However, one can argue that, despite this mathematical equivalence, quantum states are not physically or conceptually equivalent to quantum probabilities, and that quantum probabilities are just inferred from the quantum states. In other words, one should distinguish the representational content of QM from its normative content or how it is used. The representational content of QM is about the relations between the observing and the observed systems in which the quantum state represents the information of the former about the latter. However, when we use the QM formalism then we make use of probabilities, and thus probabilities only reflect its normative content. The objection is well taken. However, on the one hand we feel that disentangling the mathematics of the theory from its physical or metaphysical content is more dangerous than useful, increasing the interpretation underdetermination; on the other hand, we still maintain that one's viewpoint on probabilities should be consistent with one's other interpretive commitments, especially on the nature of the quantum state and the observing system, even if just for the sake of conceptual economy. Put otherwise, the interpretation of QM comes in a package deal.

### 3.3. The two horns of the interpretation underdetermination

Through our analysis of the pragmatist epistemic interpretations we have arrived at two interpretive tendencies: one in terms of agency and experience exemplified in its most radical form in QBism and the other in terms of purely physical, experimental context, as in the original Bohr's interpretation or generic interactions between any two physical systems, as in RQM. The crucial issue regarding the interpretation underdetermination of QM, within this epistemic-pragmatist class, concerns the difference between an ‘agent’ and a ‘context’ which in turn relates to the difference between an ‘experience’ and a generic physical ‘interaction’. This difference rests, from a more general philosophical perspective, also on the view one holds on what constitutes a "law of nature", i.e. whether it corresponds to something 'out there' in the world that we discover or something that we, as cognitive agents, 'construct'.

Can the agent be contextualized and thereby reduced to a context? Is an experience just an interaction? If one tries to explain everything in terms of the context of an experiment, then what one is left with is just an interaction between two systems. Moreover, by contextualizing every attribute related to agency in an experiment one can give a purely objective account of physical experiments. This accords with a dispositionalist interpretation of probability. There is no room for ‘experience’ in this picture. On the contrary, when agents interact with the world they 'experience' the world. Yet it seems that there is an irreducible subjective aspect to experience which cannot be contextualized or 'objectified'. And this requires a subjective interpretation of probability, although possibly constrained in a number of ways, and thus made 'objective' to some degree.

Moreover, an agent is an entity that uses the record of its past interactions with the world to modify and guide her future behavior or action. In other words, an agent uses the physical regularities, principles and relations expressed in the laws of nature to guide her future interaction with the world. So, the definition of agency closely relates to our interpretation of laws of nature. If laws are just representations of the structure of the world, then the relational character of QM (assuming one accepts it) simply reflects a relational structure of the world: even an electron can be regarded as an "observer" which "assigns" a quantum state to another electron, and it is thus an "agent", in the sense that it interacts with the other electron, its past history of interactions with it can be encoded in the quantum state, and, in conjunction with the laws of QM, it suffice to compute the results of its future interactions with it. This is fine as far as QM is concerned (in the sense that holding this view does not contradict anything in QM) but results in a rather empty notion of agency, that would apply to any physical system. On the other hand, if laws, including those expressed by QM, have an epistemic and normative character, intrinsically referring to an agent that constructs and uses them, to modify and guide her future interactions with the world, this requires a richer notion of agency. Regarding probabilities, in the representational picture, a propensity or dispositionalist interpretation of probability in terms of the context of experiment or purely physical aspects of an interaction is appropriate. On the other hand, in the normative picture, laws of nature have epistemic and normative character; the notions of agency and/or subjectivity are above and over the context of the experiment or of the physical interaction; this picture calls for a subjective interpretation of probability. This, we believe, is the gist of the two horns of the interpretation underdeterminaton of epistemic-pragmatist interpretations of QM.

The issue can be resolved in two ways within QM: a) in terms of no-go theorems that show that a refined notion of "observer", one that makes it more of an "agent", is required by the formalism itself (or, on the contrary, results that somehow show that such requirements are not needed or even contradict some aspect of QM); we discussed this possibility above; b) by some mathematically and conceptually compelling derivation of QM from information-theoretic, probabilistic or physical postulates which explicitly refer to agency in its richer form (or, viceversa, that entirely avoid any such reference). There has been a wave of such reconstructions in recent years aiming at deriving the quantum mechanical formalism from information-theoretic or probabilistic axioms (e.g. Hardy 2011; Masanes and Mueller 2011; Appleby et al. 2017; Höhn 2017). Of course, these axiomatic reconstructions can not rule out any of the existing pragmatist epistemic interpretations (or other). However, they can give us reasons for preferring one of the interpretations over the others, if a reconstruction can be more naturally formulated within the framework of it, for example because the chosen axioms can be argued to be the precise formulation of some of its basic tenets. In fact, that this is possible can be gauged from the fact that some of these approaches have been first proposed within (and in the form of) attempts to derive QM axiomatically (see for example RQM or Mueller's, but probably also Brukner-Zeilinger and QBism). In other words, axiomatic reconstructions based on one specific perspective may give it a decisive heuristic advantage over the others. In more detail, when QM is situated within a formal framework it is subjected to a kind of 'theoretical experiment' (Janas et al. 2022). Each particular framework imposes constraints on the nature of quantum systems and these constraints inform us about what QM is and the behavior of systems within that framework. To give a concrete example, the derivation of the Born rule for quantum probabilities (i.e. the specific generalization of classical probability theory encoded in QM) from the basic principles on which a given interpretation is based would represent a strong argument to prefer it over the others.

Otherwise, the underdetermination can only be resolved (to the extent in which purely conceptual or interpretative issues can be resolved at all) outside of QM; this should then happen, we argue, at the level of philosophical analysis of notions like 'agency', 'probability' and 'law of nature'. Although it is true that arguments about laws of nature would probably not break the tie explicitly or directly, the point is that the general view one takes about laws of nature (e.g. ontic vs epistemic, representational vs normative, etc) can give indirect weight to different perspectives on the realism that is compatible with QM and the role and nature of the context/observer, which is, as we also argued, a point of difference between different approaches we discussed. So, strong philosophical support for one or the other view of these three (related) notions will translate into stronger philosophical support of one or the other of the QM interpretations we discussed in this contribution.

## 4. Conclusion

In this paper, we have investigated the similarities and differences between the main epistemic-pragmatist interpretations of QM. It was argued that these different QM interpretations can be regarded as different points within a space that can be delineated in terms of a set of core claims which are emphasized to varying degrees (and in some cases not at all), and we take to define the epistemic-pragmatist class (beyond the acceptance of the QM formalism as it is, without calling for its completion or modification). We have identified these core commitments to be: 1) the epistemic (or anyway not ontic) view of quantum states; 2) a metaphysics of participatory realism; 3) an epistemology of perspectival objectivity. All these core commitments point to new, radical and, in our opinion, extremely interesting philosophical perspectives that deserve, and in fact require, further analysis. We leave this analysis, however, for future work. Given this shared core, one can translate between these QM interpretations by emphasizing or deemphasizing in each interpretation some aspects of the formulation of the same basic commitments, or accepting a stronger or weaker definition of the terms entering their formulation. Because of this translatability, we face to some extent an interpretational underdetermination. We argued however that key differences remain between the various QM interpretations in this class and that they have to do prominently with the notion of "observer", or more generally with the nature of the two systems entering in the relations encoded in quantum states, and with how one understands the probabilities encoded or computed from the same quantum states. We have also argued that, staying within QM, one can only resolve this underdetermination thanks to results constraining what physical system can play the role of observer in quantum relations, or by new axiomatic formulations of QM itself. Lacking that, the only resolution of the underdetermination can be based on extra-quantum-theoretical considerations, i.e. philosophical arguments. The relevant philosophical issues, we claim, are those concerning the nature of agency, probabilities and physical laws. If one broadens their viewpoint to this higher level of analysis then one has the resources to break the interpretation underdetermination. Carrying out this higher level philosophical analysis is, again, work for another day.